\documentclass{article}[12pt]
\usepackage{graphicx}
\usepackage{hyperref}
\usepackage{amssymb}
\usepackage{multirow}
\usepackage{caption}
\usepackage{subcaption}
\usepackage[left=1.5cm,right=1.5cm,top=1.5cm,bottom=1.5cm]{geometry}
\usepackage{authblk}
\usepackage{amsmath}
\begin{document}
\title{Interacting quark matter and $f(Q)$ gravity: A new paradigm in exploring the properties of quark stars}
\author[1]{Debadri Bhattacharjee\thanks{debadriwork@gmail.com}}
\author[2]{Koushik Ballav Goswami\thanks{koushik.kbg@gmail.com}}
\author[3]{Pradip Kumar Chattopadhyay\thanks{pkc$_{-}76$@rediffmail.com}}
\affil[1,2,3]{IUCAA Centre for Astronomy Research and Development (ICARD), Department of Physics, Cooch Behar Panchanan Barma University, Vivekananda Street, District: Cooch Behar, \\ Pin: 736101, West Bengal, India}
\maketitle
\begin{abstract}{Perturbative Quantum Chromodynamics corrections and the colour superconductivity indicate that strongly interacting matter can manifest unique physical behaviours under extreme conditions. Motivated by this notion, we investigate the interior structure and properties of quark stars composed of interacting quark matter, which provides a comprehensive avenue to explore the strong interaction effects, within the framework of $f(Q)$ gravity. A unified equation of state is formulated to describe various phases of quark matter, including up-down quark matter $(2SC)$, strange quark matter $(2SC+s)$, and the Colour-Flavor Locked $(CFL)$ phase. By employing a systematic reparametrisation and rescaling, the number of degrees of freedom in the equation of state is significantly reduced. Utilising the Buchdahl-I metric ansatz and a linear $f(Q)$ functional form, $f(Q)=\alpha_{0}+\alpha_{1}Q$, we derive the exact solutions of the Einstein field equations in presence of the unified interacting quark matter equation of state. For the $2SC$ phase, we examine the properties of quark stars composed of up-down quark matter. For the $(2SC+s)$and $CFL$ phases, we incorporate the effects of a finite strange quark mass $(m_{s}\neq0)$. The Tolman-Oppenheimer-Volkoff equations are numerically solved to determine the maximum mass-radius relations for each phase. Our results indicate that the model satisfies key physical criteria, including causality, energy conditions, and stability requirements, ensuring the viability of the configurations. Furthermore, the predicted radii for certain compact star candidates align well with observational data. The study highlights that quark stars composed of interacting quark matter within the $f(Q)$ gravity framework provide a robust and physically consistent stellar model across all considered phases.}
\end{abstract}
\section{Introduction}\label{sec1} 
Einstein's General Theory of Relativity (GR) has revolutionised our understanding of gravitational phenomena, from the solar system to cosmological scales \cite{Will}. Verified experimentally by Arthur Eddington in 1919, GR remains the most successful framework for explaining gravitational interactions. Its predictions have transformed ultra-dense compact objects, such as black holes (BH), neutron stars (NS), and white dwarfs (WD), from theoretical concepts into observable entities, particularly after the 1967 discovery of pulsars.

Recent advances in observational capabilities have greatly enhanced our understanding of high-energy phenomena at smaller cosmic scales, fostering significant interest in the behaviour of matter under the extreme densities and temperatures found in NS. Experimental replication of these conditions remains infeasible, making the equation of state (EoS) for dense matter heavily reliant on theoretical models \cite{Baym,Feroci}. Considerable efforts have been directed toward addressing this challenge, particularly by exploring the fundamental composition of compact stars. Proposed forms of exotic matter include Bose-Einstein condensates, quark-gluon plasma, and high-temperature superconductors. For NS and quark stars, an accurate EoS for strongly interacting matter at high densities and low temperatures is vital, theoretically derivable from Quantum Chromodynamics (QCD). However, the non-perturbative nature of QCD at finite baryon densities and the lattice QCD "sign problem" complicate first-principles solutions \cite{Forcrand}. At lower densities, nuclear interactions dominate and are modeled using relativistic or non-relativistic effective theories \cite{Lattimer}. However, in the ultra-dense stellar core, densities surpass nuclear saturation, necessitating descriptions in terms of deconfined quarks. Due to the absence of more realistic models, this high-density regime often relies on simplified models like the MIT bag model \cite{Farhi}, which incorporates a bag constant to encapsulate interaction effects. Itoh \cite{Itoh} first introduced the idea that NS interior may have sufficient housing conditions of quark matter cores. Pioneering work by Madsen \cite{Madsen} established the significance of strange quarks for stability in compact stars, leading to the concept of Strange Stars (SS) or Strange Quark Stars (SQS) \cite{Madsen,Baym1,Alcock,Glendenning}, where Strange Quark Matter (SQM) is hypothesised as the QCD ground state \cite{Witten}. 

However, the recent monumental work of Holdom et al. \cite{Holdom} demonstrated that if the flavour dependent feedback of gaseous quark assembly on QCD vacuum is considered, then, the bulk energy per baryon of quark matter composed of only $u$ and $d$ quarks (udQM) reduces in comparison to SQM. Hence, udQM can be relatively more stable compared SQM. Moreover, considering the finite size effects, the udQM can be termed the ground state of baryonic matter for sufficiently large baryon number density \cite{Holdom}. Interestingly, Zhang \cite{Zhang} has studied the udQM via gravitational waves and proposed a new interpretation for the binary merger event GW170817 where he showed that $u$ and $d$ quark stars (udQS) may be one of the components of the detected binary system, at least. Based on this fundamental principle, a number of phenomenological and experimental models have been postulated \cite{Zhang,Wang,Zhao,Xia,Acharya,Piotrowski}. Now, in the pursuit of a more realistic EoS, the strong interaction among the fundamental building blocks of matter must be taken into consideration. According to QCD, this fundamental interaction is largely influenced by the interactions among quarks and gluons. Consequently, QCD provides a comprehensive theoretical framework for understanding the behaviour of matter under extreme conditions, such as those found in the early universe or within SQS. In this context, the Interacting Quark Matter (IQM) has emerged as a prime candidate in describing the interior of SQS considering the impact of strong interaction and the vicinity of hyperdense compact cores. In light of the recent LIGO-VIRGO observations, many studies have shed light on the properties of SQS \cite{Zhou,Burgio,Horvath,Roupas,Harko} and udQS \cite{Ren,Zhang1,Cao} involving the presence of IQM. IQM accounts for the interquark effects arising from perturbative quantum chromodynamics (pQCD) and incorporates aspects of colour superconductivity. Now, the corrections due to pQCD are induced through the gluon mediated interactions \cite{Farhi,Fraga,Fraga1}. Further, in the quark matter, the condensation of spin-0 Cooper pair, which are antisymmetric in the flavour space, gives rise to the colour superconductivity \cite{Alford,Rajagopal,Lugones}. This can lead to the formation of two-flavour colour superconductivity, where $u$ and $d$ quarks form pairing interactions, or to a colour-flavour-locked $(CFL)$ phase, where $u$, $d$ and $s$ quarks pair with one another in an antisymmetric configuration. The pairing of $u$ and $d$ is termed $2SC$ phase and if $s$ quarks are introduced into the group, it is termed $2SC+s$ phase. Existing models of non-interacting SQM and udQM, with their current bag constant values, are inadequate to explain the unusually high mass of the observed lighter companion object of GW190814 \cite{Abbott}. This discrepancy has prompted interest in investigating whether strongly interacting quark stars (IQSs), composed of IQM, could fulfil these mass constraints and offer a viable model for the lighter companion's characteristics. Using this notion of IQM in IQS, researchers have investigated the possible formation of compact stars in different aspects of stellar modeling \cite{Pretel,Errehymy1,Tangphati}.  

In this manuscript, we explore the possibility of IQS in the framework of $f(Q)$ gravity. Shortly after GR was introduced, Weyl \cite{Weyl} made the first modifications by incorporating higher-order invariants into the Einstein-Hilbert action, aiming to unify electromagnetism and gravitation. Over time, two primary geometrical modifications to GR emerged: extensions based on curvature-such as $f(R),~f(T),~f(R,T)$ theories-and extensions based on torsion and non-metricity, like $f(Q)$ gravity. Theories based on torsion and non-metricity \cite{Einstein}, equivalent to GR, include the Teleparallel Equivalent of General Relativity (TEGR) \cite{Hayashi,Sauer} and the Symmetric Teleparallel Equivalent of General Relativity (STEGR) \cite{Nester}. In TEGR, gravity in flat spacetime with torsion is formulated using tetrads and spin connections as key variables. By constraining curvature and non-metricity tensors to zero, the Weitzenb\"ock connection can be adopted, leaving tetrads as fundamental entities, and making this choice a gauge selection that does not alter the physical interpretation of the theory. In STEGR, however, gravity is characterised by non-metricity rather than curvature or torsion, with the `coincidence gauge' setting the metric tensor as the primary variable.

An extension of STEGR, $f(Q)$ gravity, parallels $f(R)$ theory and has been widely applied \cite{Jimenez,Heisenberg}. Hohmann et al. \cite{Hohmann} studied gravitational wave polarization and propagation in Minkowski space within $f(Q)$ gravity, while Soudi et al. \cite{Soudi} found that gravitational wave polarization significantly impacts strong-field gravity. $f(Q)$ theory has been applied to late-time cosmic acceleration \cite{Lazkoz}, black holes \cite{Ambrosio,Calza}, bouncing cosmologies \cite{Bajardi}, and observational constraints \cite{Khyllep,Ayuso}. Comprehensive reviews are available in several works \cite{Heisenberg2,Barros,Jimenez2,Anagnostopoulos,Flathmann,Ambrosio2}. Recently, the $f(Q)$ theory has gained considerable attention in astrophysical research, offering new perspectives on compact stellar structures. Adeel et al. \cite{Adeel} presented a comprehensive approach to study the physical properties of compact stars in $f(Q)$ gravity. Maurya et al. \cite{Maurya2} applied gravitational decoupling within $f(Q)$ gravity to explore mass and radius constraints for the lighter component of GW190814 and other self-bound strange stars. Errehymy et al. \cite{Errehymy} proposed a singularity-free charged strange star model within $f(Q)$ gravity, while Lohakare et al. \cite{Lohakare} analysed the impact of gravitational decoupling on the maximum mass and stability of strange stars under linear $f(Q)$ action. Additional studies have also incorporated $f(Q)$ theory in analysing stable stellar configurations \cite{Bhar,Bhar1,Bhar2,Gul}.

For a more accurate physical model, it is crucial to treat the interior fluid of dense compact objects as anisotropic. Ruderman \cite{Ruderman} and Canuto \cite{Canuto} emphasised that anisotropic pressures may arise in compact stars at densities exceeding nuclear levels, with type 3A superfluidity potentially influencing the core dynamics \cite{Kippenhahn}. Quantum mechanically, this superfluidity results from Cooper pair formation at low temperatures, leading to collective behaviors akin to He-3 superfluids \cite{Broglia}. Anisotropy may also originate from fermionic fields, electromagnetic interactions, pion condensation, or viscosity effects \cite{Sawyer,Sawyer1,Herrera}. Bowers and Liang \cite{Bowers} introduced pressure anisotropy to analyse compactness and mass-radius relations, while Heintzmann and Hillebrandt \cite{Heintzmann} showed that with sufficient anisotropy, mass and redshift limits remain unrestricted. Several studies \cite{Carter,Maurya,Maurya1,Deb,Mak,Mak1,Hernandez,Bhattacharjee} have explored the impact of pressure anisotropy on the gross properties of relativistic stellar modeling.

Recent observations of massive pulsars and the quest for understanding extreme gravity regimes motivate us for exploring alternative theories of gravity. $f(Q)$ gravity, a modification of GR, offers a promising avenue to study compact objects like NS and SQS. Using an IQM EoS, which accounts for strong interactions between quarks at high densities, allows for a more realistic description of the stellar matter. By modeling stellar configurations in $f(Q)$ gravity with this EoS, we can investigate how modified gravity affects the properties of compact objects, potentially leading to new insights into the nature of gravity in extreme environments and the composition of dense matter. Recently, several studies have focused on maximum mass and radius relations of compact objects using the IQM EoS and anisotropy ansatz \cite{Pretel,Errehymy1,Tangphati}. However, this is the first manuscript that presents a novel exact solution to the Einstein field equations (EFE) within the framework of $f(Q)$ gravity, uncovering intriguing aspects of the interior structure of quark stars.

The rest of the paper is organised in the following way: section~\ref{sec2} describes the properties of interacting quark matter and a parametrised unified quark matter EoS is established here. In section~\ref{sec3}, we restrict the choice of parameter space through the evaluation of energy per baryon for different choices of colour superconductivity $(\Delta)$, strange quark mass $(m_{s})$ and bag constant $(B_{g})$. The mathematical fundamentals of $f(Q)$ gravity as expressed in section~\ref{sec4} and the exact solution of EFE in presence of IQM EoS and $f(Q)$ gravity is obtained in section~\ref{sec5}. In GR, the constants of the model are evaluated by matching the interior solution and exterior vacuum Schwarzschild space-time at the stellar boundary. However, in presence of modified theories of gravity, this boundary conditions are subjected to modifications introduced by the concerned theories of gravity. Such modifications and the resulting boundary conditions are addressed in section~\ref{sec6}. Section~\ref{sec7} demonstrates the determination of maximum mass and radius through the numerical solution of TOV equations for the $2SC$, $2SC+s$ and $CFL$ phases. Additionally, we have provided predictions for the radii of few compact stars in this section. The physical validity of the proposed model, across the three phases, is established through the radial behaviour of key parameters, alongside verification of the causality condition and essential energy criteria, as discussed in section~\ref{sec8}. Within the parameter space, the stability of the present model is assessed in section~\ref{sec9}. Finally, we summarise the main findings of this study in section~\ref{sec10}.

\section{Properties of Interacting Quark Matter: Thermodynamics and Equation of State}\label{sec2}
Following the work of Alford and Rajagopal \cite{Alford1} and accounting for the pQCD corrections, we express the general form of the free energy $(\Omega)$ of superconducting quark matter as \cite{Alford2,Weissenborn}:
\begin{equation}
	\Omega=-\frac{\xi_{4}}{4\pi^{2}}\mu^{4}+\frac{\xi_{4}(1-a_{4})}{4\pi^{2}}\mu^{4}-\frac{\xi_{2a}\Delta^{2}-\xi_{2b}m_{s}^{2}}{\pi^{2}}\mu^{2}-\frac{\mu_{e}^{4}}{12\pi^{2}}+B_{g}, \label{eq1}
\end{equation}
where, $\mu$ and $\mu_{e}$ denote that average chemical potentials contributed by quarks and electrons respectively. In eq.~\eqref{eq1}, the first term is contributed by the distribution of unpaired free quark gas, the second term arises from the pQCD, specifically representing the effects of one-gluon exchange in gluon interactions up to the order of $\mathcal{O}(\alpha_{s}^{2})$. Additionally, from a phenomenological perspective, the variation of $a_{4}$ from $a_{4}=1$ to very small values accounts for the higher order contributions \cite{Alford2,Weissenborn}. Notably, $a_{4}=1$ corresponds to a vanishing pQCD corrections. In the third term, the presence of $m_{s}$ takes care of the finite value of strange quark mass and $\Delta$ represents the colour-superconductivity effects. In the last term, $B_{g}$ is the bag constant which denotes the effective difference between the perturbative and non-perturbative vacuum energies and the size of $B_{g}$ may depend on the choice of flavour \cite{Holdom}. According to the work of Zhang and Mann \cite{Zhang1}, we note that depending on the numerical values of the constant coefficients present in eq.~\eqref{eq1}, we obtain three different phases for the different types of quark matter present in the configuration, ${\it viz.}$, 
\begin{equation}
	(\xi_{4},\xi_{2a},\xi_{2b})=
	\begin{cases}
		\Bigg(\Big[(\frac{1}{3})^{\frac{4}{3}}+(\frac{2}{3})^{\frac{4}{3}}\Big]^{-3},1,0\Bigg)~~~~~\text{2SC Phase}\\
		(3,1,\frac{3}{4})~~~~~~~~~~~~~~~~~~~~~~~~~~~\text{2SC+s Phase}\\
		(3,3,\frac{3}{4})~~~~~~~~~~~~~~~~~~~~~~~~~~~\text{CFL Phase}
	\end{cases}
	\label{eq2}
\end{equation}
Now, the thermodynamic variables, such as energy density, pressure and quark number density can be extracted from the free energy $(\Omega)$ expressed in eq.~\eqref{eq1} through the relations given below:
\begin{equation}
	p=-\Omega,~~~~~~~~n_{q}=-\frac{\partial\Omega}{\partial\mu},~~~~~~~~~~~n_{e}=-\frac{\partial\Omega}{\partial\mu_{e}},~~~~~~~~~~~ \rho=\Omega+n_{q}\mu+n_{e}\mu_{e}, \label{eq3}
\end{equation}   
Moreover, to narrow down the parameter space, we define a new parameter $\eta$ as
\begin{equation}
	\eta=\frac{\xi_{2a}\Delta^{2}-\xi_{2b}m_{s}^{2}}{\sqrt{\xi_{4}a_{4}}}. \label{eq4}
\end{equation}
Here, $\eta$ characterises the strength of the strong interaction \cite{Zhang1}. Using eq.~\eqref{eq3}, we obtain
\begin{equation}
	n_{q}=\frac{\xi_{4}a_{4}}{\pi^{2}}\mu^{3}+\frac{\eta\sqrt{\xi_{4}a_{4}}}{\pi^{2}}2\mu, ~~~~~~~~~~~~~~~~n_{e}=\frac{\mu_{e}^{3}}{3\pi^{2}}, \label{eq5}
\end{equation}
and, 
\begin{equation}
	\rho=\frac{3\xi_{4}a_{4}}{4\pi^{2}}\mu^{4}+\frac{\mu_{e}^{4}}{4\pi^{2}}+\frac{\eta\sqrt{\xi_{4}a_{4}}}{\pi^{2}}\mu^{2}+B_{g}. \label{eq6}
\end{equation}
Using eq.~\eqref{eq1} and eq.~\eqref{eq6}, we get the EoS in the form
\begin{equation}
	p=\frac{1}{3}(\rho-4B_{g})+\frac{4\eta^{2}}{9\pi^{2}}\Bigg(-1+\sqrt{1+3\pi^{2}\frac{(\rho-B_{g})}{\eta^{2}}}\Bigg). \label{eq7}
\end{equation}
However, in the present formalism, to obtain a tractable set of relativistic solutions of the Einstein field equation we parametrise the EoS in a linearised form as
\begin{equation}
	p=\frac{1}{3}(\rho-4B_{eff}), \label{eq8}
\end{equation}
where, $B_{eff}=B_{g}-\frac{3(\xi_{2a}\Delta^{2}-\xi_{2b}m_{s}^{2})}{4\pi^{2}}\mu^{2}+\frac{\eta\sqrt{\xi_{4}a_{4}}}{4\pi^{2}}\mu^{2}$. In the present context, the general linearised form of EoS provided above unifies the $2SC$, $2SC+s$ and $CFL$ phases. Hence, we denote eq.~\eqref{eq8} as the unified IQM EoS. Interestingly, for $\eta\rightarrow0$ in eq.~\eqref{eq7}, we obtain the conventional non-interacting EoS for quark matter, whereas, for the other extreme end of limit, i.e, for $\eta\rightarrow\infty$, eq.~\eqref{eq7} proceeds towards a special form as given below:
\begin{equation}
	p=\rho-2B_{g}. \label{eq9}
\end{equation}  
From eq.~\eqref{eq9}, we note that, the inclusion of strong interaction reduces the surface mass density of quarks from $4B_{g}$, for non-interacting quark systems, to $2B_{g}$ in case of IQS. Moreover, the sound velocity for quark matter increases from $\frac{dp}{d\rho}=\frac{1}{3}$ up to $1$. 
\section{Constraining the choice of $\Delta$ and $m_{s}$ for different phases}\label{sec3} One way to characterise the necessary constraints, on the permissible numerical values of colour superconductivity $(\Delta)$ and strange quark mass $(m_{s})$ for the three different phases, is through the evaluation of energy per baryon number $(\mathcal{E_{B}})$. The energy per baryon $(\mathcal{E_{B}})$ of the most stable nuclei, i.e., $^{56}Fe$ is $930.4~MeV$. Hence, for the quark assembly in $2SC$, $2SC+s$, and, $CFL$ phases, if the values of $\mathcal{E_{B}}$ are less than $930.4~MeV$, the configuration can be termed stable. In this section, we have studied the variation of $\mathcal{E_{B}}$ with respect to $\Delta$ for different choices of $m_{s}$ and a bag constant $B_{g}=70~MeV/fm^{3}$ and the results are graphically represented. 
\begin{figure}[h]
	\centering
	\includegraphics[width=8cm]{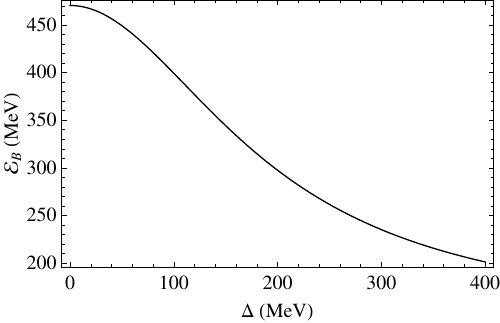}
	\caption{Energy per baryon, $(\mathcal{E_{B}})$ vs. colour superconductivity, $(\Delta)$ for $2SC$ phase}
	\label{fig1}
\end{figure}
\begin{figure}[h]
	\begin{subfigure}{.33\textwidth}
		\hspace{-1cm}
		\includegraphics[width=8cm]{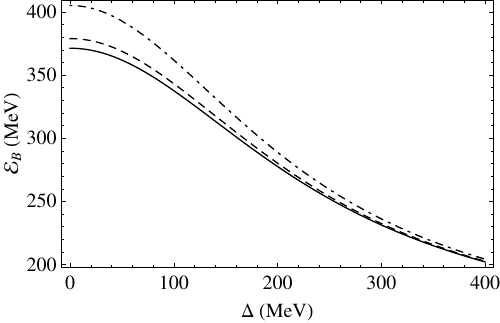}
		\caption{}
		\label{fig2a}
	\end{subfigure}%	
	\hfill
	\begin{subfigure}{.33\textwidth}
		\hspace{-2cm}
		\includegraphics[width=8cm]{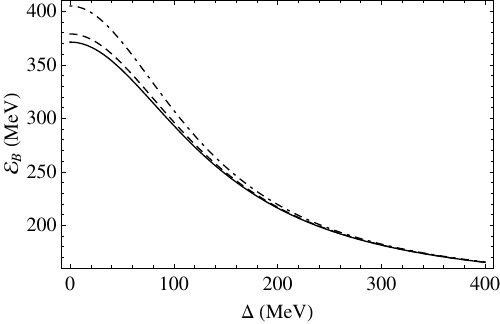}
		\caption{}
		\label{fig2b}
	\end{subfigure}
	\caption{Energy per baryon, $(\mathcal{E_{B}})$ vs. colour superconductivity, $(\Delta)$ for (a) $2SC+s$ phase and (b) $CFL$ phase. Here, the solid, dashed and dotdashed lines represent $m_{s}=0,~50$ and $100~MeV$ respectively.}
	\label{fig2}
\end{figure}
Figure~\ref{fig1}, represents the variation of energy per baryon $(\mathcal{E_{B}})$ with respect to colour superconductivity $(\Delta)$ in $2SC$ phase. It is observed that the udQM actually possesses a lower energy per baryon number in comparison to SQM, which aids to the higher stability of udQM. Further, from figure~\ref{fig2}, we note the influence of strange quark mass $(m_{s})$ on $\mathcal{E_{B}}$ for $2SC+s$ and $CFL$ phases respectively. Figure~\ref{fig2} demonstrates that for particular value of $\Delta$, $\mathcal{E_{B}}$ increases with increasing value of $m_{s}$. However, the numerical values of $\mathcal{E_{B}}$ for the maximum value of $m_{s}$ considered in this manuscript, i.e., $m_{s}=100~MeV$, remains well within the stable SQM range for $2SC+s$ and $CFL$ phases respectively. Therefore, from figures~\ref{fig1} and \ref{fig2}, we note that the numerical choices for strange quark mass $(m_{s})$, bag constant $B_{g}=70~MeV/fm^{3}$ and a superconductivity parameter, $\Delta=100~MeV$, are suitable in the present formalism. Further, we note that for the above parametric choice of $\Delta$ and $B_{g}$, the maximum allowed value of strange quark mass $(m_{s})$ that ensures $\mathcal{E_{B}}<930.4~MeV$ are $262~MeV$ and $308~MeV$ for $2SC+s$ and $CFL$ phases respectively. 
\section{Fundamentals of $f(Q)$ gravity}\label{sec4}
In the framework of symmetric teleparallel $f(Q)$ gravity, the action integral is represented as follows:
\begin{equation}
	\mathfrak{S}=\int \sqrt{-g}d^{4}x\Bigg[\frac{1}{2}f(Q)+\lambda^{kij}_{l}R^{l}_{kij}+\tau^{ij}_{k}T^{k}_{ij}+\mathfrak{L_{m}}\Bigg], \label{eq10}
\end{equation}
Here, $g$ represents the determinant of the fundamental metric tensor $g_{ij}$, i.e., $g=|g_{ij}|$. The function $f(Q)$ relates to non-metricity $Q$. The terms $\lambda^{kij}_{l}$ and $\tau^{ij}_{k}$ are Lagrangian multipliers, $R^{l}_{kij}$ denotes the Riemann tensor, and $T^{k}_{ij}$ is the torsion tensor. The Lagrangian densities for matter is described by $\mathfrak{L_{m}}$. In terms of the affine connection $(\Gamma^{k}_{ij})$, the non-metricity is expressed as:
\begin{equation}
	Q_{kij}=\nabla_{k}g_{ij}=\delta_{k}g_{ij}-\Gamma^{l}_{ij}g_{ij}-\Gamma^{l}_{ik}g_{jl}. \label{eq11}
\end{equation}
Here, $\nabla_{k}$ denotes the covariant derivative. Additionally, the affine connection can be decomposed into three components as follows:
\begin{equation}
	\Gamma^{k}_{ij}=\epsilon^{k}_{ij}+K^{k}_{ij}+L^{k}_{ij}. \label{eq12} 
\end{equation}
In eq.~\eqref{eq12}, 
\begin{itemize}
	\item The Levi-Civita connection, denoted as $\epsilon^{k}_{ij}$, is defined in terms of the fundamental metric tensor $g_{ij}$ as follows:
	\begin{equation}
		\epsilon^{k}_{ij}=\frac{1}{2}g^{kl}\Big(\partial_{i}g_{lj}+\partial_{j}g_{il}-\partial_{l}g_{ij}\Big). \label{eq13}
	\end{equation}
	\item The contorsion, described by $K^{k}_{ij}$, is expressed as:
	\begin{equation}
		K^{k}_{ij}=\frac{1}{2}T^{k}_{ij}+T_{(_{i}~k~_{j})}. \label{eq14}
	\end{equation}
	In the context of STEGR, the contorsion $K^{k}_{ij}$ corresponds to the antisymmetric component of the affine connection, which is mathematically expressed as: $T^{k}_{ij}=2\Gamma^{k}_{[ij]}=\Gamma^{k}_{ij}-\Gamma^{k}_{ji}$.
	\item The term $L^{k}_{ij}$ expresses the deformation and is written as:
	\begin{equation}
		L^{k}_{ij}=\frac{1}{2}Q^{k}_{ij}+Q_{(_{i}~k~_{j})}. \label{eq15}
	\end{equation}
\end{itemize}
The superpotential related to non-metricity is defined as
\begin{equation}
	P^{kij}=-\frac{1}{4}Q^{kij}+\frac{1}{2}Q^{(ij)k}+\frac{1}{4}(Q^{k}-\tilde{Q}^{k})g^{ij}-\frac{1}{4}\delta^{k(_{i}Q_{j})}, \label{eq16}
\end{equation}
where, $Q^{k}$ and $\tilde{Q}^{k}$ represent the independent traces of $Q_{kij}$ and are written as:
\begin{equation}
	Q_{k}\equiv {Q_{k}}^{i}_{i}, ~~~~~~~~~~~ \tilde{Q}^{k}=Q_{i}^{ki}. \label{eq17}
\end{equation} 
Hence, the non-metricity scalar can be formulated as:
\begin{equation}
	Q=-g^{ij}\Big(L^{k}_{lj}L^{l}_{ik}-L^{l}_{il}L^{k}_{ij}\Big)=-Q_{kij}P^{kij}. \label{eq18}
\end{equation}
Now, we derive the gravitational field equations by varying the Einstein-Hilbert action, as expressed in eq.~\eqref{eq10}, with respect to the metric tensor $g_{ij}$, yielding the following expression:
\begin{equation}
	\frac{2}{\sqrt{-g}}\nabla_{k}\Big(\sqrt{-g}f_{Q}P^{k}_{ij}\Big)+\frac{1}{2}g_{ij}f+f_{Q}\Big(P_{ikl}Q^{kl}_{j}-2Q_{kli}P^{kl}_{j}\Big)=-T_{ij}, \label{eq19}
\end{equation}
where, $f_{Q}=\frac{\partial f}{\partial Q}$ and $T_{ij}$ represents the energy-momentum tensor associated to the contribution of matter distribution. In a more general manner, we can describe the energy-momentum $(T_{ij})$ is the form:
\begin{equation}
	T_{ij}=-\frac{2}{\sqrt{-g}}\frac{\delta(\sqrt{-g}\mathfrak{L_{m}})}{\delta g^{ij}}. \label{eq20}
\end{equation}
\section{Exact solution of EFE in the framework of f(Q) gravity with IQM EoS}\label{sec5}
To reiterate, this study explores a novel generalised approach to modeling quark stars within the framework of $f(Q)$ gravity. Now, the static spherically symmetric spacetime, a cornerstone assumption in gravitational theories, provides significant insights into diverse astrophysical phenomena. Accordingly, a static spherically symmetric spacetime has been adopted for the analysis in this work. In the static spherically symmetric curvature coordinate system, $(t,r,\theta,\phi)$, the line element is represented in the form:
\begin{equation}
	ds^2=-e^{2\nu(r)}dt^2+e^{2\lambda(r)}dr^2+r^2(d\theta^2+sin^2\theta d\phi^2). \label{eq21}
\end{equation}
Substituting eq.~\eqref{eq21} in eq.~\eqref{eq18}, we obtain the following expression of the non-metricity as given below:
\begin{equation}
	Q=-\frac{2e^{-2\lambda(r)}}{r}\Big(2\nu'(r)+\frac{1}{r}\Big), \label{eq22}
\end{equation}
Here, the prime indicates differentiation with respect to the radial coordinate $r$. Notably, $Q$ depends on a mathematical framework where parallel transport within the 
$Q$-geometry involves neither rotations nor distortions. In essence, $Q$ is constructed based on the vanishing coefficients of the affine connections. Further, the anisotropic perfect fluid distribution is defined as:
\begin{equation}
	T_{ij}=(\rho+p_{t})u_{i}u_{j}+p_{t}g_{ij}+(p_{r}-p_{t})v_{i}v_{j}, \label{eq23}
\end{equation}
Here, $u_{i}$ denotes the four-velocity of the anisotropic perfect fluid, while 
$v_{i}$ represents the spacelike unit vector oriented along the radial direction. These quantities satisfy the conditions $u^{i}u_{i}=-1$ and $v^{i}v_{i}=1$. In Eq.~(\ref{eq23}), $\rho$, $p_{r}$ and $p_{t}$ correspond to the energy density, radial pressure, and tangential pressure of the fluid, respectively. By combining the equation of motion from Eq.(\ref{eq19}) with the anisotropic perfect fluid distribution described in Eq.(\ref{eq23}), we derive a generalised set of non-zero components of the Einstein field equations in the following form:
\begin{eqnarray}
	\frac{f(Q)}{2}-f_{Q}\Big[Q+\frac{1}{r^{2}}+\frac{2e^{-2\lambda}}{r}(\nu'+\lambda')\Big]=8\pi\rho, \label{eq24} \\
	-\frac{f(Q)}{2}+f_{Q}\Big[Q+\frac{1}{r^{2}}\Big]=8\pi p_{r}, \label{eq25} \\
	-\frac{f(Q)}{2}+f_{Q}\Big[\frac{Q}{2}-e^{-2\lambda}\Big\{\nu''+2~\Big(\frac{\nu'}{2}+\frac{1}{2r}\Big)(\nu'-\lambda')\Big\}\Big]=8\pi p_{t}, \label{eq26} \\
	\frac{cot\theta}{2}Q'f_{QQ}=0, \label{eq27}
\end{eqnarray}
where the prime (') denotes differentiation with respect to $r$. Utilising eq.~\eqref{eq27}, we obtain the form of the linear $f(Q)$ action, which is expressed as:
\begin{equation}
	f(Q)=\alpha_{0}+\alpha_{1}Q, \label{eq28}
\end{equation}
Here, $\alpha_{0}$ carries the dimension of $Km^{-2}$, while $\alpha_{1}$ is a dimensionless parameter. Utilising eqs.~\eqref{eq24}, \eqref{eq25}, \eqref{eq26}, \eqref{eq27}, and \eqref{eq28}, the field equations within the framework of the modified $f(Q)$ gravity are simplified to an exact set of expressions as follows:
\begin{eqnarray}
	\frac{1}{2r^{2}}\Big[r^{2}\alpha_{0}-2\alpha_{1}e^{-2\lambda}(2r\lambda'-1)-2\alpha_{1}\Big]=8\pi\rho, \label{eq29}\\
	\frac{1}{2r^{2}}\Big[-r^{2}\alpha_{0}-2\alpha_{1}e^{-2\lambda}(2r\nu'+1)+2\alpha_{1}\Big]=8\pi p_{r}, \label{eq30} \\
	\frac{e^{-2\lambda}}{2r}\Big[-r\alpha_{0}e^{2\lambda}-2r\alpha_{1}\nu''-2\alpha_{1}(r\nu'+1)(\nu'-\lambda')\Big]=8\pi p_{t}. \label{eq31}
\end{eqnarray} 
To derive the exact set of solutions within this formalism, we apply the Buchdahl-I ansatz \cite{Buchdahl}, which encompasses nearly all solutions of the static EFE coupled with a perfect fluid distribution. This ansatz is expressed as follows:
\begin{equation}
	e^{2\lambda}=\frac{2(1+\chi r^{2})}{2-\chi r^{2}}, \label{eq32}
\end{equation} 
Substituting the metric ansatz, as given in eq.~\eqref{eq32}, in eq.~\eqref{eq29}, we obtain the analytical expression of the energy density $(\rho)$ as:
\begin{equation}
	\rho= \frac{\alpha_{0}+2\alpha_{0}\chi r^{2}-9\alpha_{1}\chi+\alpha_{0}\chi^{2}r^{4}-3\alpha_{1}\chi^{2}r^{2}}{16\pi(1+\chi r^{2})^{2}}. \label{eq33}
\end{equation}
Now, employing eq.~\eqref{eq33} in the generalised IQM EoS, expressed in eq.~\eqref{eq8}, we obtain the expression for radial pressure $p_{r}$ in the form:
\begin{equation}
	p_{r}=\frac{1}{3}\Bigg( \frac{\alpha_{0}+2\alpha_{0}\chi r^{2}-9\alpha_{1}\chi+\alpha_{0}\chi^{2}r^{4}-3\alpha_{1}\chi^{2}r^{2}}{16\pi(1+\chi r^{2})^{2}}-4B_{g}\Bigg). \label{eq34}
\end{equation}
Substituting eq.~\eqref{eq34} in eq.~\eqref{eq30}, we obtain the metric potential of the $g_{tt}(=e^{2\nu})$ component in the following form:
\begin{equation}
	\nu=\frac{\Bigg(\Big(6\alpha_{0}-7\alpha_{1}\chi-96\pi B_{g}\Big)\log{[\chi r^{2}-2]}+\chi\Big(2\alpha_{0}r^{2}-32\pi B_{g}r^{2}+6\alpha_{1}+\alpha_{1}\log{[1+\chi r^{2}]}\Big)\Bigg)}{6\alpha_{1}\chi}. \label{eq35}
\end{equation}
Now, employing eqs.~\eqref{eq32} and \eqref{eq35} in eq.~\eqref{eq31}, we get the tangential pressure $(p_{t})$ in the form:
\begin{equation}
	p_{t}=\frac{1}{144\pi\alpha_{1}(\chi r^{2}-2)(1+\chi r^{2})^{3}}\Big(4\alpha_{0}^{2}+\chi^{4}r^{10}+6\alpha_{1}(9\alpha_{1}\chi-\alpha_{0})+f_{1}+f_{2}+f_{3}+f_{4}+f_{5}+f_{6}\Big), \label{eq36}
\end{equation}
where,  \vspace{0.2cm}
$f_{1}=1024\pi^{2}B_{g}^{2}r^{2}(1+\chi r^{2})^{4}$, \\ \vspace{0.2cm} $f_{2}=\alpha_{0}\chi^{3}r^{8}(16\alpha_{0}-21\alpha_{1}\chi)$,\\  \vspace{0.2cm}
$f_{3}=3r^{6}\chi^{2}(8\alpha_{0}^{2}-33\alpha_{0}\alpha_{1}\chi+12\alpha_{1}^{2}\chi^{2})$,\\  \vspace{0.2cm}
$f_{4}=r^{2}(4\alpha_{0}^{2}-69\alpha_{0}\alpha_{1}\chi+117\alpha_{1}^{2}\chi^{2}$,\\  \vspace{0.2cm}
$f_{5}=r^{4}\chi(16\alpha_{0}^{2}-141\alpha_{0}\alpha_{1}\chi+171\alpha_{1}^{2}\chi^{2})$,\\  \vspace{0.2cm}
$f_{6}=-32\pi B_{g}(1+\chi r^{2})^{2}(-12\alpha_{1}+4r^{6}\alpha_{0}\chi^{2}+r^{2}(4\alpha_{0}-33\alpha_{1}\chi)+2r^{4}\chi(4\alpha_{0}-3\alpha_{1}\chi))$. \\ 
The pressure anisotropy, i.e., the difference between radial and tangential pressures is expressed as:
\begin{equation}
	\Delta_{a}=p_{t}-p_{r}. \label{eq37}
\end{equation}
The total active gravitational mass enclosed within a spherical volume of radius $R$ is given by,
\begin{equation}
	m(r)=4\pi\int_{0}^{R}\rho r^{2}~dr. \label{eq38}
\end{equation}
\section{Boundary conditions}\label{sec6}
To proceed, let us impose the boundary condition necessary for determining the constants and physical parameters to analyse the properties of the compact star. This involves ensuring a smooth matching of the interior spacetime to an appropriate exterior vacuum solution at the surface of vanishing pressure (i.e., at $r=R$). Specifically, for the functional form of $f(Q)=\alpha_{0}+\alpha_{1}Q$, we follow the approach outlined by Wang et al.\cite{Wang1}. For the off-diagonal component given in eq.~\eqref{eq23}, the solutions in the $f(Q)$ gravity framework are constrained to two specific cases: 
\begin{eqnarray} 
	f_{QQ}=0, \Rightarrow f(Q)=\alpha_{0}+\alpha_{1}Q, \label{eq38a}\\ 
	Q'=0, \Rightarrow Q=Q_{0}, \label{eq38b}
\end{eqnarray}
where $\alpha_{0}$, $\alpha_{1}$, and $Q_{0}$ are constants. It is important to observe that these results parallel those obtained in $f(T)$ gravity as discussed by B\"ohmer et al. \cite{Bohmer}, where the constraints $f_{TT}=0$ or $T'=T_{0}$ were employed for static, spherically symmetric configurations under a diagonal tetrad formalism. Moreover, if the cosmological constant is defined as $\frac{\alpha_{0}}{\alpha_{1}}$, the first solution in eq.~\eqref{eq38a} effectively reduces to GR, as it aligns with symmetric teleparallel general relativity. 

In the case of a vacuum, the physical quantities $\rho$, $p_{r}$, and $p_{t}$ all vanish. Consequently, the equations of motion, eqs.~\eqref{eq22}, \eqref{eq24}, \eqref{eq25}, \eqref{eq26}, in light of eq.~\eqref{eq38a}, reduce to the following forms:
\begin{equation} 
	\lambda'(r)+\nu'(r)=0, \label{eq38c}
\end{equation} 
\begin{equation} 
	Q=\frac{\alpha_{0}}{\alpha_{1}}-\frac{2}{r^{2}}. \label{eq38d} 
\end{equation}
From eq.~\eqref{eq38c}, we obtain: 
\begin{equation} 
	\nu(r)=-\lambda(r)+\lambda_{0}, \label{eq38e}
\end{equation} 
where $\lambda_{0}$ is an integration constant that can be absorbed into the solution by rescaling the time coordinate from $t$ to $e^{-\lambda_{0}}t$. As a result, the components $g_{rr}$ and $g_{tt}$ become reciprocal to one another, as noted by Wang et al. \cite{Wang1}. Therefore, eq.~\eqref{eq38e} simplifies to:
\begin{equation} 
	\nu(r)=-\lambda(r). \label{eq38f} 
\end{equation}
Using eqs.~\eqref{eq21}, \eqref{eq38c} and \eqref{eq38d}, the expression for the exterior $g_{rr}$ component is derived as: 
\begin{equation} 
	e^{2\lambda}=\left(1-\frac{c_{1}}{r}-\frac{\alpha_{0}}{6\alpha_{1}}r^{2}\right)^{-1}, \label{eq38g} 
\end{equation} 
where $c_{1}$ is an integration constant, and its sign is chosen for convenience. Similarly, using eqs.~\eqref{eq38f} and \ref{eq38g}, the $g_{tt}$ component is obtained as: 
\begin{equation} 
	e^{2\nu}=\Bigg(1-\frac{c_{1}}{r}-\frac{\alpha_{0}}{6\alpha_{1}}r^{2}\Bigg). \label{eq38h} 
\end{equation}
Thus, the line element for the spherically symmetric vacuum exterior spacetime, within the framework of $f(Q)$ theory of gravity, describing a compact stellar configuration can be expressed as:
\begin{equation} 
	ds^{2}=\Bigg(1-\frac{c_{1}}{r}-\frac{\alpha_{0}}{6\alpha_{1}}r^{2}\Bigg)dt^{2}+\Bigg(1-\frac{c_{1}}{r}-\frac{\alpha_{0}}{6\alpha_{1}}r^{2}\Bigg)^{-1}dr^{2}+r^{2}(d\theta^{2}+\sin^{2}\theta d\phi^{2}). \label{eq38i} 
\end{equation}
Analysing the metric in Equation~(\ref{eq38i}) reveals three key features: (i) when $c_{1}=2M$, and (ii) when the cosmological constant $\Lambda=\frac{\alpha_{0}}{\alpha_{1}}$, the metric reduces to the Schwarzschild-(anti) de Sitter solution: 
\begin{equation} 
	ds^{2}=\Bigg(1-\frac{2M}{r}-\frac{1}{3}\Lambda r^{2}\Bigg)dt^{2} + \Bigg(1-\frac{2M}{r}-\frac{1}{3}\Lambda r^{2}\Bigg)^{-1}dr^{2}+ r^{2}(d\theta^{2}+\sin^{2}\theta d\phi^{2}), \label{eq38j} 
\end{equation} 
where $M$ represents the total mass of the stellar configuration with radius $R$. In the absence of the cosmological constant, the metric reduces further to the Schwarzschild solution \cite{Schwarzschild}: 
\begin{equation} 
	ds^{2}=-\Bigg(1-\frac{2M}{r}\Bigg)dt^{2}+\Bigg(1-\frac{2M}{r}\Bigg)^{-1}dr^{2}+r^{2}(d\theta^{2}+\sin^{2}\theta d\phi^{2}). \label{eq38k} 
\end{equation}
Therefore, the observed correspondence between eqs.~\eqref{eq38i} and \eqref{eq38j} strongly suggests that such solutions are attainable only under linear form of $f(Q)$ gravity. This linearity condition effectively reduces the theory to a form equivalent GR. Again, following Wang et al. \cite{Wang1}, we note that, when a non-trivial functional form of $f(Q)$ is considered, the exact Schwarzschild solutions are not possible.

In modified theories of gravity, the equations of motion prescribed by GR are altered, necessitating corresponding modifications to the boundary conditions typically applied to stellar systems in GR \cite{Rosa}. Within the 4-dimensional spacetime manifold $(\Sigma)$, there are two distinct regions: the interior and exterior spacetimes. These regions are separated by a 3-dimensional hypersurface $(\Xi)$, which is characterised by the induced metric $h_{ij}$ in the coordinate system $X^{i}$, with indices excluding the direction orthogonal to $\Xi$. The projection tensors and the normal vector are expressed as $e^{a}_{i}=\frac{\partial x^{a}}{\partial X^{i}}$ and $n_{a}=\epsilon\delta_{a}\ell$, where $\ell$ represents the affine connection in the direction perpendicular to $\Xi$, and $\epsilon$ takes values of $-1,~0$ or $1$ for the respective timelike, null, and spacelike geodesics. Given this setup, we have $n^{a}e^{i}_{a}=0$. The induced metric $h_{ij}$ and the extrinsic curvature tensor $K_{ij}$ at the hypersurface are then written as:
\begin{equation}
	h_{ij}=e^{a}_{i}e^{b}_{j}g_{ab}, ~~~~~~~~~~~~~~~~~~~~~~K_{ij}=e^{a}_{i}e^{b}\nabla_{a}n_{b}. \label{eq39}
\end{equation} 
In the distribution formalism introduced by Rosa et al. \cite{Rosa1}, the condition $[h_{ij}=0]$ is imposed, ensuring the continuity of the induced metric at the hypersurface $\Xi$. To determine the constants in the metric potentials, it is essential to also require the continuity of the extrinsic curvature tensor at the junction. In the absence of a thin shell, this continuity condition is expressed as $[K_{ij}=0]$ \cite{Rosa1}. To calculate the extrinsic curvature, we start by considering the vacuum exterior solution obtained in eq.~\eqref{eq38k}. Notably, in the parametric space of spherically symmetric spacetime, $K_{ij}^{\pm}$ consists of only two components: $K_{00}$ and $K_{\theta\theta}=K_{\phi\phi}sin^{2}\theta$, where $(+)$ and $(-)$ signs correspond to the exterior and interior spacetimes, respectively. Using Eq.~\eqref{eq39}, we calculate the extrinsic curvature tensors for the interior and exterior spacetimes. By matching these tensors at the stellar boundary $(r=R)$, we derive the following result:
\begin{equation}
	\chi=-\frac{4M}{R^{2}(4M-3R)}. \label{eq41}
\end{equation} 
Moreover, another essential condition is that the radial pressure vanishes at the surface of the star, i.e.,
\begin{equation}
	p_{r}(r=R)=0. \label{eq42}
\end{equation}
Eq.~\eqref{eq42} is employed to derive the expression for stellar radius in the present model. However, to simplify the presentation and avoid complex mathematical expressions, we have instead tabulated the numerical values of the radii for the $2SC$, $2SC+s$ and $CFL$ phases for the compact star under consideration.  
\section{Determination of mass-radius relation from TOV equation}\label{sec7}
To obtain the maximum mass and the associated radius in presence of IQM EoS expressed in eq.~\eqref{eq8}, we have numerically solved the Tolman-Oppenheimer-Volkoff equations \cite{Tolman,Oppenheimer}, considering bag constant, $B_{g}=70~MeV/fm^{3}$, colour superconductivity, $\Delta=100~MeV$ and different choices of strange quark mass, {\it viz.}, $m_{s}=0,~50$ and $100~MeV$ for $2SC$, $2SC+s$ and $CFL$ phases respectively. We illustrate the concerned results below.
\begin{figure}[h]
	\centering
	\includegraphics[width=8cm]{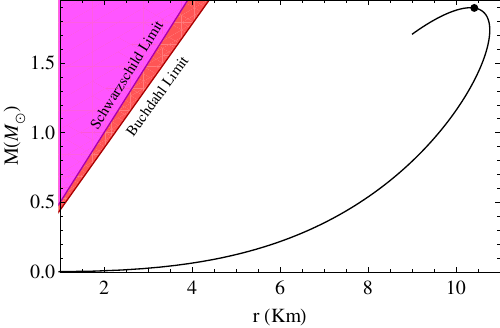}
	\caption{Mass-radius diagram for $2SC$ phase.}
	\label{fig3}
\end{figure}
\begin{figure}[h]
	\begin{subfigure}{.33\textwidth}
		\hspace{-1cm}
		\includegraphics[width=8cm]{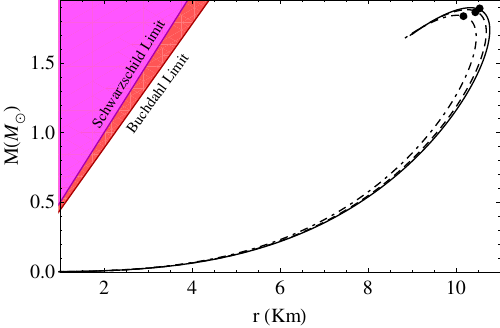}
		\caption{}
		\label{fig4a}
	\end{subfigure}%	
	\hfill
	\begin{subfigure}{.33\textwidth}
		\hspace{-2cm}
		\includegraphics[width=8cm]{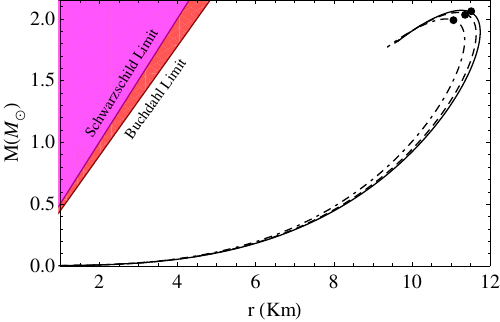}
		\caption{}
		\label{fig4b}
	\end{subfigure}
	\caption{Mass-radius diagram for (a) $2SC+s$ phase and (b) $CFL$ phase. Here, the solid, dashed and dotdashed lines represent $m_{s}=0,~50$ and $100~MeV$ respectively.}
	\label{fig4}
\end{figure}
Figure~\ref{fig3} demonstrates the maximum mass-radius plot for udQM matter in $2SC$ phase. Whereas, from figure~\ref{fig4}, we note that with increasing strange quark mass $(m_{s})$, both the maximum mass and radius decrease for the $2SC+s$ and $CFL$ phases respectively. This can be explained by the fact that a higher $m_{s}$ leads to an increase in the associated Fermi energy, as the Fermi energy of degenerate fermions depends on both the mass and Fermi momentum of the strange quark. Consequently, the EoS becomes softer thereby decreasing the maximum mass further. Additionally, a higher strange quark mass lowers the quark number density for a given pressure, as more massive quarks require greater energy. This results in fewer quarks being accommodated in a given volume at fixed pressure, causing a rapid rise in energy density at higher baryon number densities. The increased energy density accelerates gravitational collapse, leading to a reduction in both the maximum mass and the corresponding radius. Thus, as $m_{s}$ increases, the EoS softens, causing a decrease in the maximum mass, as illustrated in Fig.~(\ref{fig4}). The maximum mass and the corresponding radii are tabulated in table~\ref{tab1}.
\begin{table}[h]
	\centering
	\begin{tabular}{|cccc|}
		\hline
		Phase & Strange quark mass $(m_{s})$ & $M_{max}~(M_{\odot})$ & $R~(Km)$\\\hline
		$2SC$ & -- & 1.89 & 10.34 \\
		\hline
		\multirow{3}{*}{$2SC+s$} & 0 & 1.89 & 10.34 \\
		& 50 & 1.88 & 10.25\\
		& 100 & 1.84 & 10.04\\ \hline
		\multirow{3}{*}{$CFL$} & 0 & 2.07 & 11.27 \\
		& 50 & 2.05 & 11.17 \\
		& 100 & 1.99 & 10.89 \\
		\hline
	\end{tabular}
	\caption{Tabulation of maximum mass and the associated radius for $2SC$, $2SC+s$ and $CFL$ phases respectively.}
	\label{tab1}
\end{table} 
Following figures~\ref{fig3} and \ref{fig4} and table~\ref{tab1}, a comparison of the three phases on the basis of mass-radius relation reveals that $CFL$ phase actually accommodates a more massive stellar structure relative to the other phases \cite{KBG}. Further, we have predicted the radii of some known compact stars in table~\ref{tab1a}. Additionally, tables~\ref{tab1aa} and \ref{tab1aaa} describe the central parameters such as central density $(\rho_{0})$, surface density $(\rho_{s})$ and central pressure $(p_{0})$ for $2SC$, $2SC+s$ and $CFL$ phases respectively.  
\begin{table}[h!]
	\centering
	\begin{tabular}{|ccccccc|}
		\hline
		\multirow{2}{*}{Compact objects} & Observed mass & Observed radius & \multicolumn{4}{c|}{Radius prediction} \\ \cline{4-7}
		& $(M_{\odot})$ & $(Km)$ & $2SC$ & $m_{s}$ & $2SC+s$ & $CFL$ \\
		&&& $(Km)$ & $(MeV)$ & $(Km)$ & $(Km)$\\
		\hline
		EXO 1745-248 \cite{Ozel} & 1.4 & 11 & 10.52 & 50 & 10.46 & 11.18 \\
		\hline
		4U 1820-30 \cite{Guver1} & $1.58^{+0.06}_{-0.06}$ & $9.1^{+0.4}_{-0.4}$ & 10.68 & 250 & 9.10 & 9.77 \\
		\hline
		LMC X-4 \cite{Rawls} & $1.04^{+0.09}_{-0.09}$ & $8.301^{+0.2}_{-0.2}$ & 9.72 & 250 & 8.64 & 9.05 \\
		\hline
		HER X-1 \cite{Abubekerov} & $0.85^{+0.15}_{-0.15}$ & $8.1^{+0.41}_{-0.41}$ & 9.17 & 200 & 8.51 & 8.93 \\
		\hline
		PSR J1903+0327 \cite{Freire} & $1.667^{+0.021}_{-0.021}$ & $9.438^{+0.03}_{-0.03}$ & 10.75 & 200 & 9.57 & 10.36 \\
		\hline
	\end{tabular}
	\caption{Radius prediction of some known compact objects from TOV mass-radius relation in $2SC$, $2SC+s$ and $CFL$ phases respectively.}
	\label{tab1a}
\end{table}
\begin{table}
	\centering
	\begin{tabular}{|c|c|c|c|}
		\hline
		\multirow{2}{*}{Compact} & \multicolumn{3}{c|}{$2SC$ phase} \\
		\cline{2-4}
		\multirow{2}{*}{objects} & $\rho_{0}\times10^{15}$ & $\rho_{s}\times10^{15}$ & $p_{0}\times10^{34}$\\
		& $(gm/cm^{3})$ & $(gm/cm^{3})$ & $(dyn/cm^{2})$ \\
		\hline
		EXO 1745-248 & 0.46 & 0.28 & 0.10\\
		4U 1820-30 & 0.52 & 0.30 & 1.81\\
		LMC X-4 & 0.47 & 0.32 & 0.48\\
		HER X-1 & 0.46 & 0.32 & 0.09\\
		PSR J1903+0327 & 0.55 & 0.30 & 2.68\\
		\hline
	\end{tabular}
	\caption{Tabulation of central parameters of few known compact objects for $2SC$ phase.}
	\label{tab1aa}
\end{table}
\begin{table}
	\centering
	\begin{tabular}{|c|c|c|c|c|c|c|}
		\hline
		\multirow{2}{*}{Compact} &\multicolumn{3}{c|}{$2SC$ phase} & \multicolumn{3}{c|}{$CFL$ phase}\\
		\cline{2-7} 
		\multirow{2}{*}{objects} & $\rho_{0}\times10^{15}$ & $\rho_{s}\times10^{15}$& $p_{0}\times10^{34}$&$\rho_{0}\times10^{15}$ & $\rho_{s}\times10^{15}$& $p_{0}\times10^{34}$\\
		& $(gm/cm^{3})$ & $(gm/cm^{3})$ & $(dyn/cm^{2})$ & $(gm/cm^{3})$ & $(gm/cm^{3})$ & $(dyn/cm^{2})$\\
		\hline
		EXO 1745-248 & 0.47 & 0.28 & 0.16 & 0.40 & 0.25 & 0.10 \\
		4U 1820-30 & 0.91 & 0.46 & 8.23 & 0.71 & 0.38 & 4.45 \\
		LMC X-4 & 0.63 & 0.41 & 0.04 & 0.56 & 0.37 & 0.09\\
		HER X-1 & 0.58 & 0.40 & 0.23 & 0.50 & 0.36 & 0.14\\
		PSR J1903+0327 & 0.82 & 0.42 & 7.60 & 0.62 & 0.34 & 3.82\\ 
		\hline
	\end{tabular}
	\caption{Tabulation of central parameters of few known compact objects for $2SC+s$ and $CFL$ phases.}
	\label{tab1aaa}
\end{table}
\newpage
\section{Physical attributes of the proposed model}\label{sec8} 
For a stellar model to be considered physically viable, it is essential that the associated physical parameters derived from the model are realistic and consistent with observable phenomena. To qualify as a compact star, the model must satisfy the fundamental conditions of regularity and physical plausibility that are inherent to relativistic stellar structures. In this pursuit, we have considered the Low-Mass X-ray Binary (LMXB) 4U 1608-52, having a mass of $1.74~M_{\odot}$ \cite {Guver}, to study the physical viability of the present model under the combined influence of interacting quark system and $f(Q)$ formalism. Moreover, following the work of Maurya et al. \cite{Maurya3}, we have considered $\alpha_{0}=10^{-46}~Km^{-2}$ and $\alpha_{1}=-0.6$ as arbitrary choices. Now, to start with, we have used eq.~\eqref{eq42} to estimate the radii of LMXB 4U 1608-52 for different values of $m_{s}$ across the three phases, within the framework of $f(Q)$ theory of gravity and the results are tabulated in table~\ref{tab2}. From table~\ref{tab2}, it is evident that as $m_{s}$ increases the radius decreases which points towards a softer EoS. Notably, in describing the physical characteristics of the present model, we have used the radii from table~\ref{tab2} for the three phases.    
\begin{table}[h]
	\centering 
	\begin{tabular}{|ccc|}
		\hline
		Phases & Strange quark mass $(m_{s})$ & Predicted radius $(Km)$ \\ 
		\hline
		$2SC$ & -- & 9.33 \\
		\hline
		\multirow{3}{*}{$2SC+s$} & 0 & 9.33 \\
		& 50 & 9.28 \\
		& 100 & 9.13 \\ 
		\hline
		\multirow{3}{*}{$CFL$} & 0 & 9.95 \\
		& 50 & 9.89 \\
		& 100 & 9.70 \\
		\hline	
	\end{tabular}
	\caption{Radius prediction for 4U 1608-52 from model.}
	\label{tab2}
\end{table}
\subsection{Radial variation of thermodynamic parameters} For a realistic stellar configuration, the behaviour of the inherent thermodynamic attributes, i.e., energy density $(\rho)$, radial $(p_{r})$ and tangential $(p_{t})$ components of pressure and pressure anisotropy $(\Delta_{a})$, must maintain certain viability conditions which are discussed below:
\begin{itemize}
	\item {\bf Energy Density:} Figures~\ref{fig5} and \ref{fig6} show the diagram of $\rho$ vs. $r$ for the three phases respectively. The energy density $(\rho)$ of udQM phase is shown in figure~\ref{fig5}. From figures~\ref{fig6a} and \ref{fig6b}, we note that as $m_{s}$ increases, the energy density $(\rho)$ of the quarks, in $2SC+s$ and $CFL$ phases, increases. The energy density reaches a peak value at the core and thereafter diminishes towards the stellar boundary which describes a physically viable stellar model.
	\begin{figure}[h!]
		\centering
		\includegraphics[width=8cm]{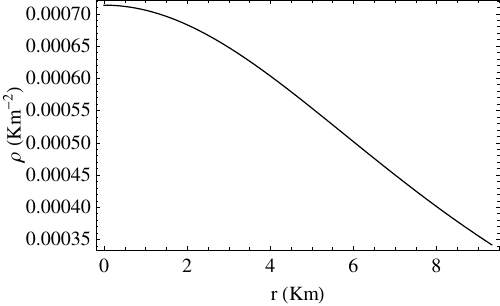}
		\caption{Radial variation of energy density $(\rho)$ for $2SC$ phase.}
		\label{fig5}
	\end{figure}
	\begin{figure}[h!]
		\begin{subfigure}{.33\textwidth}
			\hspace{-1cm}
			\includegraphics[width=8cm]{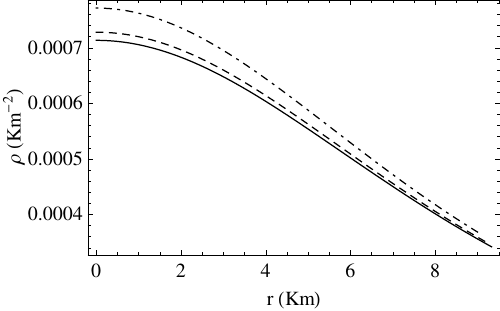}
			\caption{}
			\label{fig6a}
		\end{subfigure}%	
		\hfill
		\begin{subfigure}{.33\textwidth}
			\hspace{-2cm}
			\includegraphics[width=8cm]{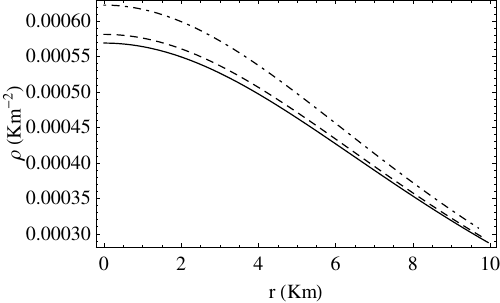}
			\caption{}
			\label{fig6b}
		\end{subfigure}
		\caption{Radial variation of energy density $(\rho)$ for (a) $2SC+s$ phase and (b) $CFL$ phase. Here, the solid, dashed and dotdashed lines represent $m_{s}=0,~50$ and $100~MeV$ respectively.}
		\label{fig6}
	\end{figure}
	\newpage
	\item {\bf Radial pressure:} The radial pressure counterbalances the inward gravitational force which prevents the collapse of compact stars under their own weight. Moreover, for a physically realistic model, the variation of the radial pressure $(p_{r})$ with respect to radial parameter $(r)$ must be monotonically decreasing from centre to the stellar surface which is depicted in figures~\ref{fig7} and \ref{fig8}. Figures~\ref{fig7}, \ref{fig8a} and \ref{fig8b} illustrate the radial pressure profiles for the $2SC$, $2SC+s$ and $CFL$ phases respectively. Evidently, from figure~\ref{fig8}, with increasing strange quark mass $(m_{s})$, the radial pressure increases. However, due to the decreasing energy density towards the boundary, the magnitude of radial pressure diminishes.    
	\begin{figure}[h!]
		\centering
		\includegraphics[width=8cm]{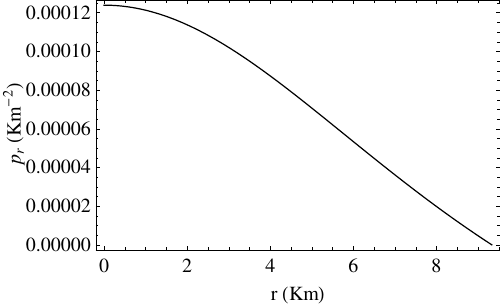}
		\caption{Radial variation of radial pressure $(p_{r})$ for $2SC$ phase.}
		\label{fig7}
	\end{figure}
	\begin{figure}[h!]
		\begin{subfigure}{.33\textwidth}
			\hspace{-1cm}
			\includegraphics[width=8cm]{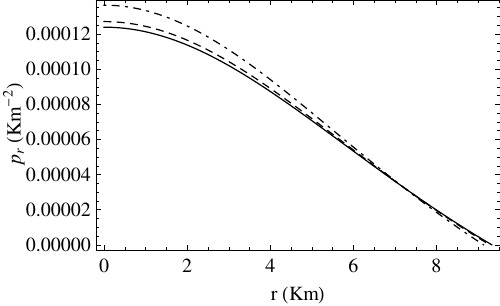}
			\caption{}
			\label{fig8a}
		\end{subfigure}%	
		\hfill
		\begin{subfigure}{.33\textwidth}
			\hspace{-2cm}
			\includegraphics[width=8cm]{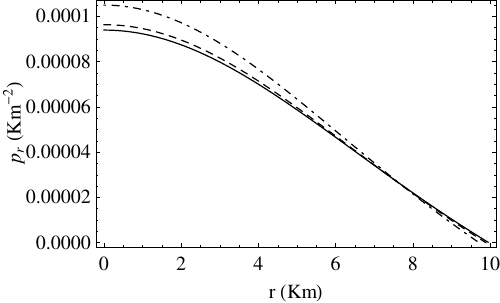}
			\caption{}
			\label{fig8b}
		\end{subfigure}
		\caption{Radial variation of radial pressure $(p_{r})$ for (a) $2SC+s$ phase and (b) $CFL$ phase. Here, the solid, dashed and dotdashed lines represent $m_{s}=0,~50$ and $100~MeV$ respectively.}
		\label{fig8}
	\end{figure}
	\newpage
	\item {\bf Tangential pressure:} Tangential pressure $(p_{t})$ actually maintains the spherical shape of the compact star by counteracting the outward radial pressure $(p_{r})$. Depending on various factors, such as specific stellar composition, EoS etc., the behaviour of $p_{t}$ can vary. However, the nature of radial and tangential pressure profiles are similar, i.e., consistently decreasing from the stellar core to boundary and it is demonstrated in figures~\ref{fig9} and \ref{fig10} for the three phases within the parameter space used here. Notably, $p_{t}$ is also positive at all internal points of the configuration, which points towards a physically acceptable stellar structure. Here, figures~\ref{fig9}, \ref{fig10a} and \ref{fig10b} describe the radial variation of tangential pressure $p_{t}$ profiles across the three phases. Moreover, we note that with increasing strange quark mass $(m_{s})$ the magnitude of $p_{t}$ increases within the choice of parameter space.   
	\begin{figure}[h!]
		\centering
		\includegraphics[width=8cm]{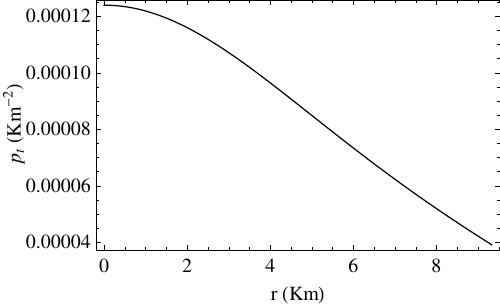}
		\caption{Radial variation of tangential pressure $(p_{t})$ for $2SC$ phase.}
		\label{fig9}
	\end{figure}
	\begin{figure}[h!]
		\begin{subfigure}{.33\textwidth}
			\hspace{-1cm}
			\includegraphics[width=8cm]{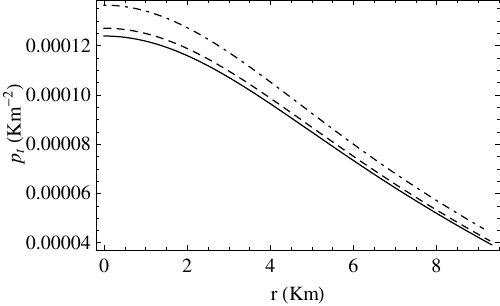}
			\caption{}
			\label{fig10a}
		\end{subfigure}%	
		\hfill
		\begin{subfigure}{.33\textwidth}
			\hspace{-2cm}
			\includegraphics[width=8cm]{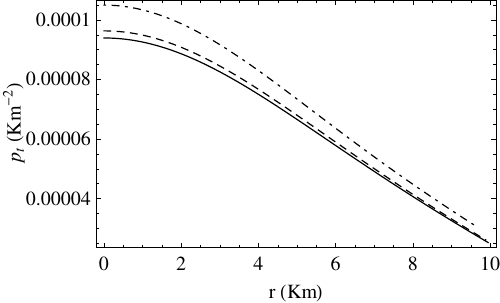}
			\caption{}
			\label{fig10b}
		\end{subfigure}
		\caption{Radial variation of tangential pressure $(p_{t})$ for (a) $2SC+s$ phase and (b) $CFL$ phase. Here, the solid, dashed and dotdashed lines represent $m_{s}=0,~50$ and $100~MeV$ respectively.}
		\label{fig10}
	\end{figure}
	\newpage
	\item {\bf Pressure anisotropy:} The anisotropy in pressure $(\Delta_{a})$ is the difference between the radial and tangential pressures, i.e., $\Delta_{a}=p_{t}-p_{r}$. Figures~\ref{fig11}, \ref{fig12a} and \ref{fig12b} illustrate the radial variation of pressure anisotropy for the three phases respectively. Additionally, we note that $\Delta_{a}>0$ in all the phases, which depict a repulsive anisotropic behaviour. Notably, from figure~\ref{fig12}, it is observed that with increasing strange quark mass $(m_{s})$, $\Delta_{a}$ increases within this parameter space. This feature can be attributed to several reasons, {\it viz.}, (i) strange quarks $(s)$ are heavier than $u$ and $d$ quarks. Now, increasing the mass of $s$ quarks leads to a mass disparity which ultimately alters the momentum distribution and the interaction among quarks. This imbalance can lead to to an uneven distribution of forces, contributing to anisotropy in the pressure, (ii) larger $m_{s}$ softens the EoS. As a result the effective pressures in different directions become more sensitive to mass difference, which amplifies the anisotropic behaviour, (iii) The interaction between quarks in the dense matter is influenced by their masses. An increase in strange quark mass strengthens the role of strong interactions and modifies the confinement properties which may lead to differential pressure contributions in different directions, thereby enhancing the pressure anisotropy.     
	\begin{figure}[h!]
		\centering
		\includegraphics[width=8cm]{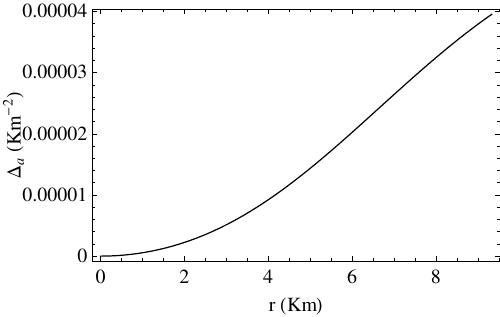}
		\caption{Radial variation of pressure anisotropy $(\Delta_{a})$ for $2SC$ phase.}
		\label{fig11}
	\end{figure}
	\begin{figure}[h!]
		\begin{subfigure}{.33\textwidth}
			\hspace{-1cm}
			\includegraphics[width=8cm]{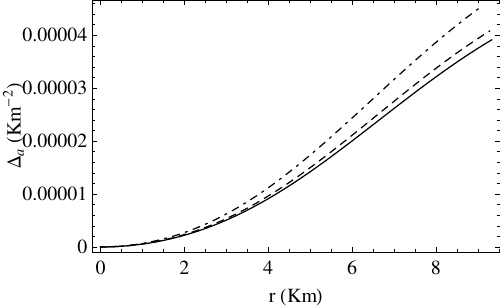}
			\caption{}
			\label{fig12a}
		\end{subfigure}%	
		\hfill
		\begin{subfigure}{.33\textwidth}
			\hspace{-2cm}
			\includegraphics[width=8cm]{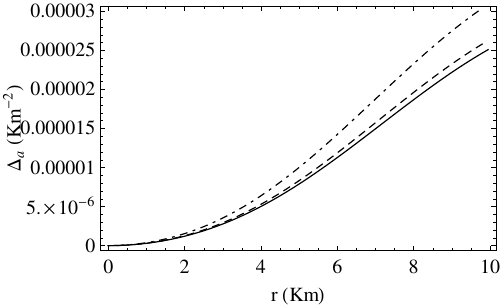}
			\caption{}
			\label{fig12b}
		\end{subfigure}
		\caption{Radial variation of pressure anisotropy $(\Delta_{a})$ for (a) $2SC+s$ phase and (b) $CFL$ phase. Here, the solid, dashed and dotdashed lines represent $m_{s}=0,~50$ and $100~MeV$ respectively.}
		\label{fig12}
	\end{figure}
\end{itemize}
\subsection{Causality condition} 
\begin{figure}[h!]
	\centering
	\includegraphics[width=8cm]{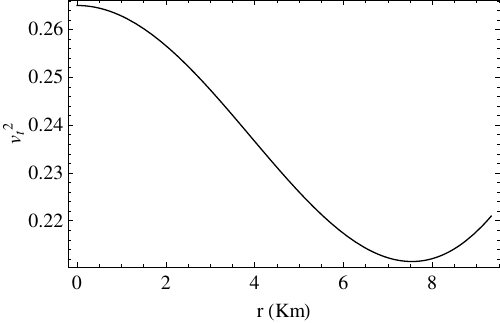}
	\caption{Radial variation of tangential sound velocity $(v_{t}^{2})$ for $2SC$ phase.}
	\label{fig13}
\end{figure}

\begin{figure}[h!]
	\begin{subfigure}{.33\textwidth}
		\hspace{-1cm}
		\includegraphics[width=8cm]{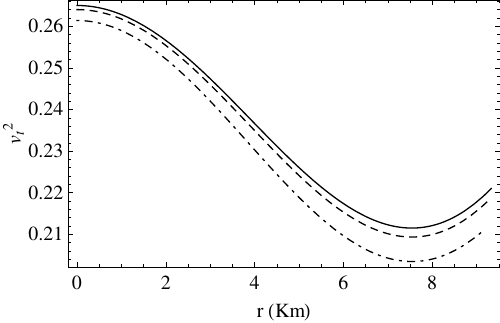}
		\caption{}
		\label{fig14a}
	\end{subfigure}%	
	\hfill
	\begin{subfigure}{.33\textwidth}
		\hspace{-2cm}
		\includegraphics[width=8cm]{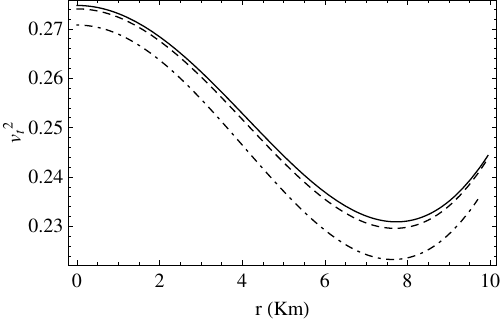}
		\caption{}
		\label{fig14b}
	\end{subfigure}
	\caption{Radial variation of tangential sound velocity $(v_{t}^{2})$ for (a) $2SC+s$ phase and (b) $CFL$ phase. Here, the solid, dashed and dotdashed lines represent $m_{s}=0,~50$ and $100~MeV$ respectively.}
	\label{fig14}
\end{figure}

To develop a realistic model for an anisotropic compact star, a crucial aspect is the analysis of sound wave propagation within its dense interior. The radial and tangential sound velocities are defined as $v_{r}^{2}=\frac{dp_{r}}{d\rho}$ and $v_{t}^{2}=\frac{dp_{t}}{d\rho}$, respectively, where $\rho$, $p_{r}$ and $p_{t}$ have been previously introduced. Adopting natural units $(\hbar=c=1)$, the causality condition enforces an upper limit on the sound velocities, requiring $v_{r}^{2}\leq1$ and $v_{t}^{2}\leq1$. Additionally, thermodynamic stability demands that $v_{r}^{2}>0$ and $v_{t}^{2}>0$. Consequently, the sound velocities must satisfy the combined constraints $0<v_{r}^{2}\leq1$ and $0<v_{t}^{2}\leq1$ throughout the stellar interior. Notably, following the IQM EoS as expressed in eq.~\eqref{eq8}, we note that the radial sound velocity, $v_{r}^{2}=\frac{dp_{r}}{d\rho}=\frac{1}{3}$, throughout all the phases, however, the tangential sound velocity $(v_{t}^{2})$ varies with strange quark mass $(m_{s})$. To address the complexity of these relationships, we have chosen to illustrate the radial variation of the tangential sound velocities graphically, as shown in figures~\ref{fig13} and \ref{fig14}. Figure~\ref{fig13} is subjected to the $2SC$ phase whereas figures~\ref{fig14a} and \ref{fig14b} denote the $2SC+s$ and $CFL$ phases respectively. Evidently, the causality conditions are consistently upheld throughout the stellar interior within the chosen parameter space.
\subsection{Energy conditions}
In gravitational theory, energy conditions impose constraints on matter distributions, serving as essential guidelines for formulating a physically valid energy-momentum tensor. These conditions offer a framework for analysing the properties of matter distributions without requiring detailed knowledge of the internal composition. Consequently, phenomena such as the formation of geometric singularities or gravitational collapse during mergers can be studied without explicitly specifying the pressure or energy density. Fundamentally, the analysis of energy conditions is an algebraic problem \cite{Kolassis}, specifically an eigenvalue problem associated with the energy-momentum tensor. In a 4-dimensional space-time, evaluating these conditions involves solving a quartic polynomial to determine its eigenvalues, which can become analytically intricate. Despite the challenges in obtaining general solutions, a physically viable fluid distribution must satisfy the dominant, strong, weak, and null energy conditions, collectively referred to as the energy conditions \cite{Kolassis,Hawking,Wald}, within the stellar configuration. In this work, following the presence of IQM EoS, we assess these energy conditions for the present stellar model using the formulations provided in \cite{Brassel, Brassel1} as: 
\begin{itemize}
	\item	NEC:~~~~$T_{ij}l^{i}l^{j}\geq0\Rightarrow\rho+p_{r}\geq0,~\rho+p_{t}\geq0$,\\
	\item	WEC:~~~~ $T_{ij}t^{i}t^{j}\geq0\Rightarrow\rho\geq0,~\rho+p_{r}\geq0,~\rho+p_{t}\geq0$,\\
	\item	SEC:~~~~$T_{ij}t^{i}t^{j}-\frac{1}{2}T_{k}^{k}t^{\sigma}t_{\sigma}\geq0\Rightarrow\rho+\sum p\geq0,~or,~\rho+p_{r}\geq0,~\rho+p_{t}\geq0,~\rho+p_{r}+2p_{t}\geq0$, \\
	\item	DEC:~~~~$T_{ij}t^{i}t^{j}\geq0\Rightarrow\rho\geq0,~\rho-p_{r}\geq0,~\rho-p_{t}\geq0$,
\end{itemize} 
where, $l^{i}$ and $t^{i}$ are time-like and null vectors respectively.
\begin{figure}[h!]
	\centering
	\includegraphics[width=8cm]{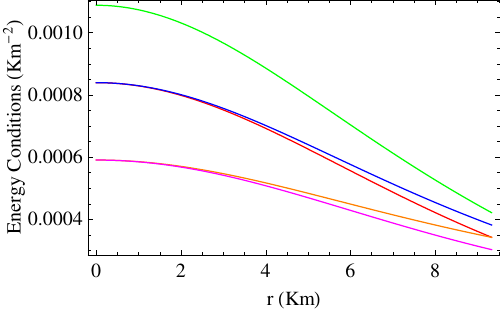}
	\caption{Radial variation of energy conditions for $2SC$ phase. Here, Red, Blue, Green, Orange and Magenta lines represent $(\rho+p_{r})$, $(\rho+p_{t})$, $(\rho+p_{r}+2p_{t})$, $(\rho-p_{r})$ and $(\rho-p_{t})$ respectively.}
	\label{fig15}
\end{figure}
\begin{figure}[h!]
	\begin{subfigure}{.33\textwidth}
		\hspace{-1cm}
		\includegraphics[width=8cm]{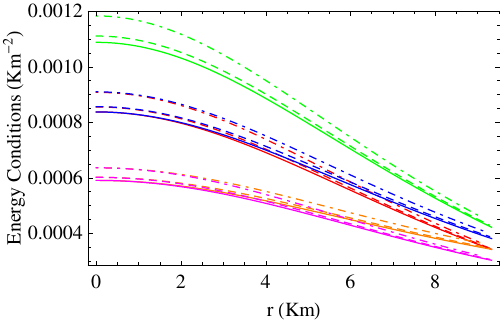}
		\caption{}
		\label{fig16a}
	\end{subfigure}%	
	\hfill
	\begin{subfigure}{.33\textwidth}
		\hspace{-2cm}
		\includegraphics[width=8cm]{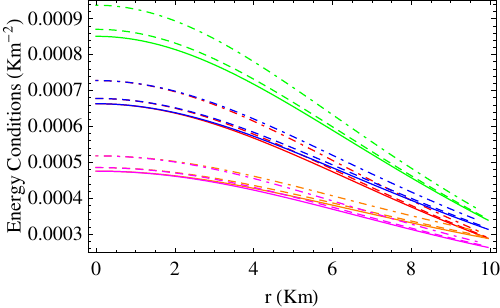}
		\caption{}
		\label{fig16b}
	\end{subfigure}
	\caption{Radial variation of energy conditions for (a) $2SC+s$ phase and (b) $CFL$ phase. Here, Red, Blue, Green, Orange and Magenta lines represent $(\rho+p_{r})$, $(\rho+p_{t})$, $(\rho+p_{r}+2p_{t})$, $(\rho-p_{r})$ and $(\rho-p_{t})$ respectively. Moreover, the solid, dashed and dotdashed lines represent $m_{s}=0,~50$ and $100~MeV$ respectively.}
	\label{fig16}
\end{figure}
Figures~\ref{fig15} and \ref{fig16} indicate that the necessary energy conditions are adequately satisfied within the parameter space utilised in the present model. 
\section{Stability analysis}\label{sec9} The stability of the proposed model is addressed through the following methods:
\begin{itemize}
	\item TOV equation in a generalised form,
	\item Herrera cracking condition, and, 
	\item Adiabatic index
\end{itemize}
\subsection{TOV equation in a generalised form} The physical viability of a stellar model, in terms of hydrostatic equilibrium, can be assessed through stability analysis, which examines the interplay of various forces within the stellar configuration. For a compact star with pressure anisotropy, this analysis considers three key force components: (i) the gravitational force $(F_{g})$, (ii) the hydrostatic force $(F_{h})$, and, (iii) the anisotropic force $(F_{a})$. To achieve stable equilibrium, these forces must collectively balance each other. In this work, we evaluate stability using the generalised form of the Tolman-Oppenheimer-Volkoff (TOV) equation \cite{Tolman, Oppenheimer}, as presented below:
\begin{equation}
	-\frac{M_{G}(\rho+p_{r})}{r^{2}}e^{\lambda-\nu}-\frac{dp_{r}}{dr}+\frac{2\Delta_{a}}{r}=0. \label{eq43}
\end{equation}
The active gravitational mass, represented as $M_{G}$ and defined in Eq.~(\ref{eq43}), can be obtained from the mass formula proposed by Tolman-Whittaker \cite{Gron}, expressed in the following form:
\begin{equation}
	M_{G}(r)=r^{2}\nu'e^{\nu-\lambda}. \label{eq44}
\end{equation}
Substituting Eq.~(\ref{eq44}) into Eq.~(\ref{eq43}), we obtain:
\begin{equation}
	-\nu'(\rho+p_{r})-\frac{dp_{r}}{dr}+\frac{2\Delta_{a}}{r}=0. \label{eq45}
\end{equation}
Using the demarcation as written below:
\begin{eqnarray}
	F_{g}=-\nu'(\rho+p_{r}), \label{eq46} \\
	F_{h}=-\frac{dp_{r}}{dr}, \label{eq47} \\
	F_{a}=\frac{2\Delta_{a}}{r}, \label{eq48}.
\end{eqnarray}
we rewrite eq.~\eqref{eq45} in the form:
\begin{equation}
	F_{g}+F_{h}+F_{a}=0. \label{eq49}
\end{equation}
By substituting eqs.~\eqref{eq33}, \eqref{eq34}, \eqref{eq35} and \eqref{eq36}, in eq.~\eqref{eq49}, the equilibrium condition outlined in the generalised TOV equations is derived. To simplify the presentation and circumvent mathematical intricacies, the stable equilibrium condition is illustrated graphically.
\begin{figure}[h!]
	\centering
	\includegraphics[width=8cm]{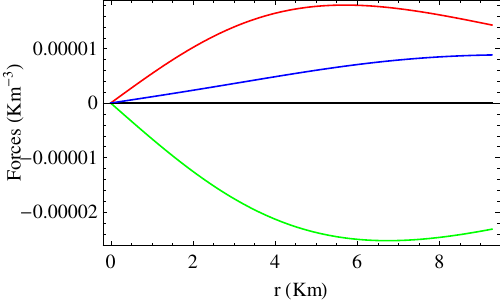}
	\caption{Radial variation of different forces for $2SC$ phase. Here, Green, Red and Blue lines represent $F_{g}$, $F_{h}$ and $F_{a}$ respectively.}
	\label{fig17}
\end{figure}
\begin{figure}[h!]
	\begin{subfigure}{.33\textwidth}
		\hspace{-1cm}
		\includegraphics[width=8cm]{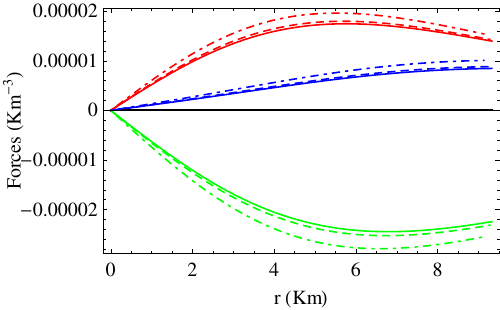}
		\caption{}
		\label{fig18a}
	\end{subfigure}%	
	\hfill
	\begin{subfigure}{.33\textwidth}
		\hspace{-2cm}
		\includegraphics[width=8cm]{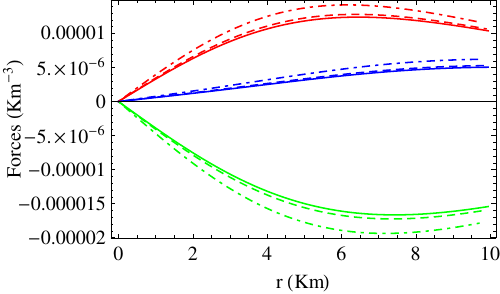}
		\caption{}
		\label{fig18b}
	\end{subfigure}
	\caption{Radial variation of different forces for (a) $2SC+s$ phase and (b) $CFL$ phase. Here, Green, Red and Blue lines represent $F_{g}$, $F_{h}$ and $F_{a}$ respectively. Moreover, the solid, dashed and dotdashed lines represent $m_{s}=0,~50$ and $100~MeV$ respectively.}
	\label{fig18}
\end{figure}
Figures~\ref{fig17} and \ref{fig18} illustrate that our mode is in stable equilibrium under the impact of difference forces, across the three phases within this parameter space.
\subsection{Herrera cracking condition} 
\begin{figure}[h!]
	\centering
	\includegraphics[width=8cm]{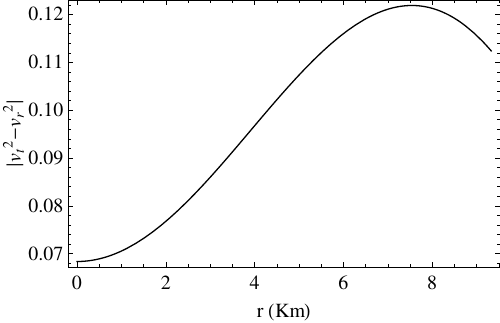}
	\caption{Radial variation of $|v_{t}^{2}-v_{r}^{2}|$ for $2SC$ phase.}
	\label{fig19}
\end{figure}

\begin{figure}[h!]
	\begin{subfigure}{.33\textwidth}
		\hspace{-1cm}
		\includegraphics[width=8cm]{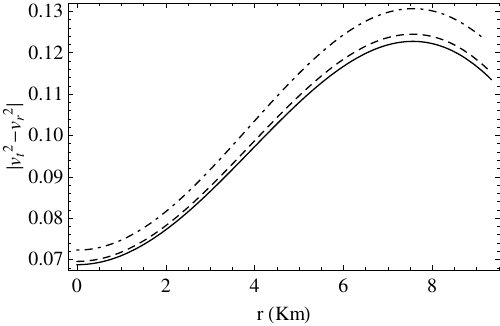}
		\caption{}
		\label{fig20a}
	\end{subfigure}%	
	\hfill
	\begin{subfigure}{.33\textwidth}
		\hspace{-2cm}
		\includegraphics[width=8cm]{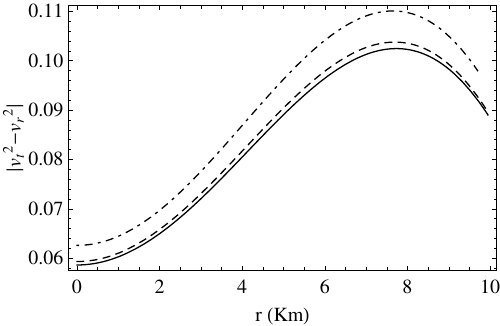}
		\caption{}
		\label{fig20b}
	\end{subfigure}
	\caption{Radial variation of $|v_{t}^{2}-v_{r}^{2}|$ for (a) $2SC+s$ phase and (b) $CFL$ phase. Here, the solid, dashed and dotdashed lines represent $m_{s}=0,~50$ and $100~MeV$ respectively.}
	\label{fig20}
\end{figure}

Herrera \cite{Herrera1} introduced the concept of "Cracking" in the context of a fluid sphere within a self-gravitating compact object. This term refers to the total radial forces of opposing signs that emerge when the system experiences perturbations. When considering the EoS, cracking can occur under two conditions: (i) in the presence of a locally anisotropic fluid or (ii) in a slowly contracting and radiating perfect fluid. Using this framework, Abreu et al. \cite{Abreu} investigated how local anisotropy influences the distribution of an anisotropic fluid. They showed that when perturbations are present, the potential stability of the system can be determined by the difference between the squares of the radial $(v_{r}^{2})$ and tangential $(v_{t}^{2})$ sound velocities, as expressed in the following relation:
\begin{equation}
	0\leq|v_{t}^{2}-v_{r}^{2}|\leq1. \label{eq50}
\end{equation}
A stellar model satisfying this relation is considered a stable structure. From figures~\ref{fig19}, \ref{fig20a} and \ref{fig20b}, it is evident that the Abreu inequality is consistently upheld throughout the stellar interior for the $2SC$, $2SC+s$ and $CFL$ phases respectively within the chosen parameter space.
\newpage
\subsection{Adiabatic index} 
\begin{figure}[h!]
	\centering
	\includegraphics[width=8cm]{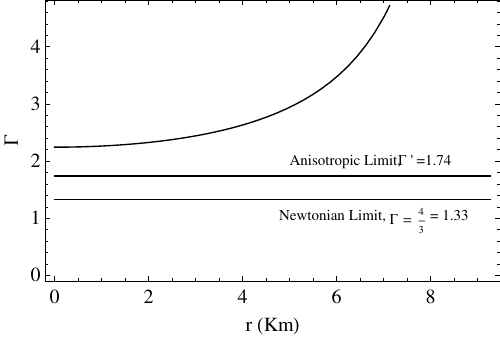}
	\caption{Radial variation of adiabatic index for $2SC$ phase.}
	\label{fig21}
\end{figure}

\begin{figure}[h!]
	\begin{subfigure}{.33\textwidth}
		\hspace{-1cm}
		\includegraphics[width=8cm]{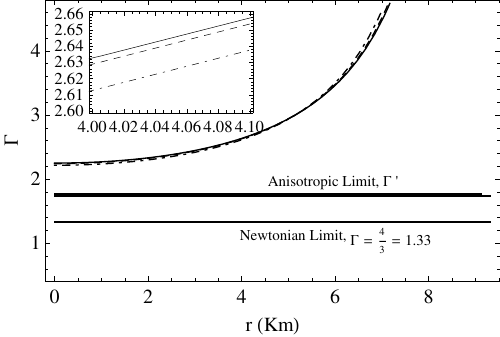}
		\caption{}
		\label{fig22a}
	\end{subfigure}%	
	\hfill
	\begin{subfigure}{.33\textwidth}
		\hspace{-2cm}
		\includegraphics[width=8cm]{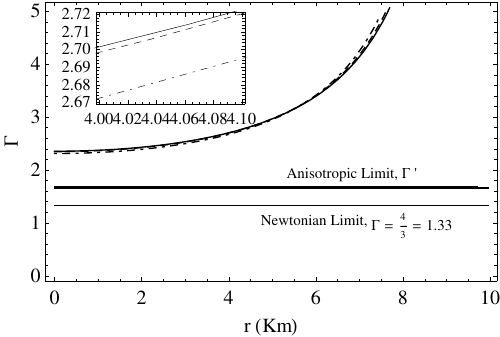}
		\caption{}
		\label{fig22b}
	\end{subfigure}
	\caption{Radial variation of adiabatic index for (a) $2SC+s$ phase and (b) $CFL$ phase. Here, the solid, dashed and dotdashed lines represent $m_{s}=0,~50$ and $100~MeV$ respectively.}
	\label{fig22}
\end{figure}

The adiabatic index $(\Gamma)$ indicates the stiffness of the EoS at a given density. In the framework of a relativistic anisotropic stellar structure, the adiabatic index is expressed as:
\begin{equation}
	\Gamma=\frac{\rho+p_{r}}{p_{r}}\frac{dp_{r}}{dr}=\frac{\rho+p_{r}}{p_{r}}v_{r}^{2}. \label{eq51}
\end{equation} 
Heintzmann and Hillebrandt \cite{Heintzmann} showed that for a Newtonian matter distribution, the adiabatic index satisfies $\Gamma>\frac{4}{3}$. Subsequently, Chan et al. \cite{Chan} extended this upper bound for the adiabatic index by taking pressure anisotropy into account, as expressed in the following form:
\begin{equation}
	\Gamma>\Gamma', \label{eq52}
\end{equation}
where, 
\begin{equation}
	\Gamma'=\frac{4}{3}-\Bigg[\frac{4}{3}\frac{(p_{r}-p_{t})}{|p'_{r}|r}\Bigg]_{max}. \label{eq53}
\end{equation}
Now, to determine the anisotropic limit as expressed in eq.~\eqref{eq53}, we have evaluated $\Gamma'$ for different values of strange quark mass $(m_{s})$ for the three phases. The results are graphically represented in figures~\ref{fig21}, \ref{fig22a} and \ref{fig22b}. 
Based on figures~\ref{fig21} and \ref{fig22}, we may assert that the current model preserves the notion of the radial variation of adiabatic index, as required for a stable anisotropic stellar model, to emerge as a physically acceptable stellar structure across the three phases, within the parameter space utilised here. 
\section{Discussion}\label{sec10}
The MIT bag model, despite its simplicity, has long been the preferred framework for studying quark matter in compact stars. Fundamental questions, such as the potential presence of deconfined matter in NS cores, have traditionally been explored while disregarding interactions among the system's constituents. Within the Standard Model of particle physics, QCD is recognised as the fundamental theory governing the strong interactions between quarks mediated by gluons. Given the QCD framework, it is evident that existing weak-coupling EoS for cold quark matter require refinement to provide a more realistic depiction of quark matter cores in compact stars.

This study adheres to a simple yet parameterised EoS for cold quark matter that addresses the limitations of the bag model and other low-energy effective models by incorporating the properties of IQM. Leveraging perturbative QCD expansions and colour superconductivity, we provide a framework that accounts for inter-quark effects arising from strong interactions, which may uncover novel physical phenomena under extreme conditions of matter and gravity subjected to the interior of compact stars. Specifically, we explore the impact of strong interactions on anisotropic quark stars composed of IQM within the symmetric teleparallel $f(Q)$ theory of gravity. By considering a spherically symmetric space-time, we employ the Buchdahl-I metric ansatz \cite{Buchdahl} and IQM EoS, as expressed in eq.~\eqref{eq8}, alongside a linear form of the $f(Q)$ action which enables the construction of singularity-free and regular solutions of the EFE under the $f(Q)$ theory of gravity. 

Based on the work of Zhang and Mann \cite{Zhang1}, it is observed that the numerical values of the constant coefficients in the EoS give rise to three distinct phases of quark matter within the configuration, {\it viz.}, (i) the pairing of $u$ and $d$ quarks (udQM) is termed the $2SC$ phase, (ii) the inclusion of $s$ quarks in udQM leads to $2SC+s$ phase and (iii) $CFL$ phase. To examine these physical acceptability of the present model in these phases, we have chosen the LMXB 4U 1608-52 \cite{Guver} as a representative candidate and analysed its properties across the three phases. Our findings, presented both numerically and graphically, focus on the effects of varying the strange quark mass $(m_{s})$ for the $2SC$, $2SC+s$ and $CFL$ phases respectively. Key characteristic features of the configuration have been identified and analysed in detail below:
\begin{itemize}
	\item We begin by imposing the restrictions on the particular choices of strange quark mass $(m_{s})$, colour superconductivity $(\Delta)$ and bag constant $(B_{g})$, through the evaluation of energy per baryon $(\mathcal{E_{B}})$ for the three phases. Figures~\ref{fig1} and \ref{fig2} illustrates the variation of $\mathcal{E_{B}}$ vs. $\Delta$ for the three phases respectively. The energy per baryon $(\mathcal{E_{B}})$ of the most stable nuclei, i.e., $^{56}Fe$ is $930.4~MeV$. Hence, to describe a stable quark matter, $\mathcal{E_{B}}$ has to be less than $930.4~MeV$. Figures~\ref{fig1} and \ref{fig2} demonstrate the variation of $\mathcal{E_{B}}$ for $B_{g}=70~MeV/fm^{3}$ and different choices of strange quark mass, $m_{s}=0,~50$ and $100~MeV$. Following figures~\ref{fig1} and \ref{fig2}, we note that the notion of $\mathcal{E_{B}}<930.4~MeV$ is well preserved in this formalism. Therefore, the numerical choices of $\Delta=100~MeV$, $B_{g}=70~MeV/fm^{3}$ and $m_{s}=0,~50$ and $100~MeV$ are justified in the present model. Moreover, from figure~\ref{fig1}, it is evident that udQM possesses a lower energy per baryon relative to SQM which further supports its stability. Moreover, with increasing $m_{s}$, we note that $\mathcal{E_{B}}$ increases for the $2SC+s$ and $CFL$ phases as shown in figures~\ref{fig2a} and \ref{fig2b} respectively.
	\item To determine the maximum mass and the corresponding radius in this interacting quark system, we have numerically solved the TOV equations within the parametric choices of microscopic parameters for the three phases. Figures.~\ref{fig3} and \ref{fig4} graphically illustrate the concerned results for the $2SC$, $2SC+s$ and $CFL$ phases respectively. Further, from figures~\ref{fig4a} and \ref{fig4b} we note that in the $2SC+s$ and $CFL$ phases,the maximum mass decreases with increasing strange quark mass $(m_{s})$, which denotes a softening of the EoS. This may be attributed to the fact that, with increasing $m_{s}$, the energy density increases which may accelerate the gravitational collapse of the structure thereby reducing the maximum mass and the associated radius. The maximum mass and radius for all the three phases are tabulated in table~\ref{tab1}. As evident from figures~\ref{fig3} and \ref{fig4} and table~\ref{tab1}, the mass-radius comparison indicates that the $CFL$ phase supports a more massive stellar structure than other phases \cite{KBG}. Moreover, in table~\ref{tab1a}, we have predicted the radii of some compact star candidates for the three phases and different values of $m_{s}$. Additionally, tables~\ref{tab1aa} and \ref{tab1aaa} describe the central density $(\rho_{0})$, surface density $(\rho_{s})$ and central pressure $(p_{0})$ across the three phases, within the parameter space used here. 
	\item The radial variation of the thermodynamic parameters such as energy density $(\rho)$, radial $(p_{r})$ and tangential $(p_{t})$ pressures are demonstrated in figures~\ref{fig5}, \ref{fig6}, \ref{fig7}, \ref{fig8}, \ref{fig9} and \ref{fig10}, throughout the three phases. We note that in all the phases $(2SC, 2SC+s$ and $CFL)$, the physical characteristic parameters maintain a monotonically decreasing nature. In particular, for the $2SC+s$ and $CFL$ phases, the energy density and pressure profiles increase with increasing value of $m_{s}$. In figures~\ref{fig11} and \ref{fig12}, the pressure anisotropy profiles are shown. Additionally, the anisotropy is positive in all the three phases, with depicts a repulsive anisotropic nature. In the context of $2SC+s$ and $CFL$ phases, we have observed that the magnitude of anisotropy increases with increasing value of $m_{s}$. Several factors contribute to this phenomenon: (i) the heavier mass of strange quarks $(s)$ compared to $u$ and $d$ quarks introduces mass disparities, affecting momentum distribution and pressure anisotropy, (ii) increased $m_{s}$ softens the EoS, amplifying directional pressure sensitivity, and (iii) stronger interactions from higher $m_{s}$ modify confinement, enhancing anisotropy. 
	\item Within the parameter space used here, the causality and the necessary energy conditions are well maintained throughout the stellar boundary.
	\item Within this IQM system and $f(Q)$ formalism, the stability of the present model is addressed through several well acknowledged methodologies, {\it viz.}, the generalised TOV equations, Hererra's concept of cracking and the radial variation of adiabatic index. Figures~\ref{fig17} and \ref{fig18} demonstrate that the present model is in hydrostatic equilibrium under the influence of different forces across the three phases. The stability criteria based on the concept of cracking and Abreu inequality \cite{Abreu} is well maintain in this model as evident from figures~\ref{fig19} and \ref{fig20}. Further, from figures~\ref{fig21} and \ref{fig22}, we have noted that the adiabatic index variation for an anisotropic stellar model is well supported in the present formalism.    
\end{itemize}
In conclusion, we may assert that the $f(Q)$ gravity framework proves effective in developing physically consistent stellar models compatible with fundamental principles in static, spherically symmetric space-times. By incorporating an IQM EoS that captures strong quark interactions at high densities, this formalism provides a realistic depiction of stellar matter under extreme conditions. Thus, the proposed model represents a stable and physically viable description to explore the structural and dynamical properties of quark stars, composed of interacting quark matter, within the $f(Q)$ gravity framework.

\section{Acknowledgments}
DB is thankful to the Department of Science and Technology (DST), Govt. of India, for providing the fellowship vide no:  DST/INSPIRE Fellowship/2021/IF210761. DB and KBG are thankful to the Department of Physics, Cooch Behar Panchanan Barma University for providing the necessary help to carry out the research work. PKC gratefully acknowledges support from IUCAA, Pune, India under Visiting Associateship programme.   

%\paragraph{Note added.} This is also a good position for notes added
%after the paper has been written.

% The bibliography will probably be heavily edited during typesetting.
% We'll parse it and, using the arxiv number or the journal data, will
% query inspire, trying to verify the data (this will probalby spot
% eventual typos) and retrive the document DOI and eventual errata.
% We however suggest to always provide author, title and journal data:
% in short all the informations that clearly identify a document.

\end{document}